\documentclass[aps,prl,twocolumn,superscriptaddress,nofootinbib]{revtex4-2}%
\pdfoutput=1 
\usepackage{CJK}
\usepackage{physics, comment}
\usepackage{amstext}
\usepackage{amsfonts,amssymb}
\usepackage{bbm}
\usepackage{graphicx}
\usepackage{float}
\usepackage{indentfirst}
\usepackage{geometry}
\usepackage{mathrsfs}
\usepackage[normalem]{ulem}
\usepackage{color}
\usepackage{amsthm}
\usepackage{graphicx}
\usepackage{dcolumn}
\usepackage{bm}
\usepackage[utf8]{inputenc}
\usepackage[T1]{fontenc}
\usepackage{booktabs, array, mathptmx, lipsum, amsmath,multirow}
\usepackage{siunitx, xcolor}
\usepackage[version=4]{mhchem}
\usepackage{thmtools}

\usepackage{mathtools}

\theoremstyle{definition}

\def\6{{\langle}}
\def\9{{\rangle}}

\newcommand{\be}{\begin{equation}}
\newcommand{\ee}{\end{equation}}
\def\gP{\mathfrak{P}}

 \newcommand{\defeq}{\vcentcolon=}

\begin{document}
\title{Unification of energy concepts in generalised phase space theories}

\author{Libo Jiang }
\email{11930020@mail.sustech.edu.cn}
\affiliation{Alumnus,
Southern University of Science and Technology (SUSTech), Shenzhen 518055, China}
\pacs{123 }
\author{Daniel R. Terno}
\email{daniel.terno@mq.edu.au}
\affiliation{Department of Physics and Astronomy, Macquarie University, Sydney, New South Wales 2109, Australia}
\pacs{23 }

\author{Oscar Dahlsten}
\email{oscar.dahlsten@cityu.edu.hk}
\affiliation{Department of Physics, City University of Hong Kong, Tat Chee Avenue, Kowloon, Hong Kong SAR, China
}%
\affiliation{Shenzhen Institute for Quantum Science and Engineering and Department of Physics,
Southern University of Science and Technology, Shenzhen 518055, China}

\affiliation{Institute of Nanoscience and Applications, Southern University of Science and Technology, Shenzhen 518055, China}
\pacs{12 }
\begin{abstract}
We consider how to describe Hamiltonian mechanics in generalised probabilistic theories with the states represented as quasi-probability distributions. We give general operational definitions of energy-related concepts. We define generalised energy eigenstates as the purest stationary states. Planck's constant plays two different roles in the framework: the phase space volume taken up by a pure state and a dynamical factor. The Hamiltonian is a linear combination of generalised energy eigenstates. This allows for a generalised Liouville time-evolution equation that applies to quantum and classical Hamiltonian mechanics and more. The approach enables a unification of quantum and classical energy concepts and a route to discussing energy in a wider set of theories.
\end{abstract}
\maketitle

\noindent{{\bf {\em Introduction.---}}}Generalised probabilistic theories (GPTs) constitute a metatheoretical framework developed within the foundations of quantum mechanics.  Two key goals of GPTs are to understand the structure of quantum theory, particularly which elements necessarily arise from its probabilistic nature, and to elucidate the relations between classical and quantum mechanics~\cite{001FivAxi,007GPT,014GPT,T:23}. Classical and quantum theories, as well as classical-quantum hybrid models, appear as special cases~\cite{014GPT,T:23,012OpeHyb}. The states in GPTs are viewed as compressed lists of probabilities of possible measurement outcomes, with specifications contingent on the system preparation and subsequent dynamical transformations. While the initial emphasis was on the underlying probabilistic structure of GPTs, there is now a growing focus on a unified treatment of energy concepts within GPTs and the exploration of natural generalisations of classical and quantum dynamics~\cite{barnum2014higher,016TunNeg,018GenHam, 022ToySHO, 022ToogeneralEOM, ZFC:05}.

In Refs.~\cite{barnum2014higher,018GenHam,016TunNeg,022ToySHO}, the quantum theory Hamiltonian was extended in various ways to GPTs.   In Ref.~\cite{016TunNeg} it was shown that the phase space formalism can represent continuous-variable generalised probabilistic theory models. Thus one may use the phase space formalism for a dynamical description of hypothetical post-quantum theories of mechanics~\cite{016TunNeg, 022ToySHO}. Using the phase space approach  Ref.~\cite{022ToogeneralEOM} recently presented post-quantum toy models of real systems including hydrogen atoms, by hypothesizing a generalised phase-space time evolution that is based on a generalization of a quantum-mechanical Moyal bracket~\cite{ZFC:05}.

These promising results create hope that we can develop self-consistent post-quantum theories of Hamiltonian mechanics as well as gain a deeper understanding of quantum and classical mechanics and their interrelations. There are certain hurdles lying ahead.
For example, the formal classical limit of quantum theory where the Planck constant is taken to zero is singular \cite{L:17}, creating a further potential block towards a unified framework:  how does one generalise this constant?  There is also the fact that well-defined energy states in classical mechanics have descriptions (for example, Liouville density written in terms of positions and momenta), that evolve explicitly, whereas quantum energy eigenstates are stationary and thus akin to functions of action and angle variables. Thus it may appear as though at least some classical energy concepts are incompatible with quantum energy concepts.

We tackle these questions via a generalised phase space approach. By introducing postulates that reduce to the standard assumptions of the quantum and classical theories in the appropriate limits, we are able to describe dynamics in terms of a generalised Hamiltonian $H(q,p)$. The generalised evolution equations are obtained with the help of a theory-specific integration kernel $\mathcal{K}(k)$. In the particular case of  $\mathcal{K}(k)\rightarrow \frac{2}{k}\delta(k-\hbar)$, the quantum evolution is recovered. When $\mathcal{K}(k)\rightarrow \frac{2}{k}\delta(k)$, the classical evolution is obtained. General functions  $\mathcal{K}(k)$ model post-quantum theories.

The Hamiltonian of a closed system is an observable with a time-invariant expectation value $\expval{E}=\int H(q,p) f(q,p) dqdp$, where $f$ is a GPT state. We construct the Hamiltonian as $H=\sum_i E_i V_{g_i} g_i$, where $E_i$ are the generalised energy eigenvalues and $g_i$ are their generalised eigenstates that are ascribed a phase space volume  $V_{g_i}$. This volume acts as a generalization of the Bohr--Sommerfeld elementary volume, reducing to $V_{g_i}=(2\pi\hbar)^n$ for a quantum system with $n$ degrees of freedom. The generalised energy eigenstates $g_i$  are defined as the purest stationary states. In other words, they are time-invariant states that are not mixtures of other time-invariant states~\cite{013GPTPha}. They coincide with the standard energy eigenstates if the GPT is a quantum theory and are uniform distributions over phase space orbits in the case of classical mechanics.

These results provide a unified framework within which one can derive statements relating to energy in such a manner that they apply directly to both quantum and classical mechanics as well as to a wider set of theories.

We proceed as follows. First, we briefly summarize the key rules of the quantum and classical phase space description. We then generalise (i) the Born rule and Planck's constant therein, (ii) the energy eigenstates, (iii) the equation of motion, and (iv) the Hamiltonian. Detailed derivations and additional results are given in an accompanying paper~\cite{longversion}.

\noindent  {{\bf \em Phase space representation.---}}   Consider first classical mechanics of a non-constrained system with a finite number of degrees of freedom. Its states and (the algebra of) observables are smooth functions on the phase
space   $\mathfrak{P}$ ~\cite{Arnold, L:17}. Mathematically,   it is a symplectic manifold that is a cotangent bundle of the configuration space with local coordinates $q$. The local coordinates on $\mathfrak{P}$ are   $z = (q, p)$   where $p$ are the generalised momenta.

The Poisson bracket of two phase space functions $\{f,g\}$  is defined as
\be
\sum_j \left(\frac{\partial f}{\partial q_j}\frac{\partial g}{\partial p_j}-\frac{\partial f}{\partial p_j}\frac{\partial g}{\partial q_j}\right)\equiv
f\big(\overset{\leftarrow}{\partial_q}\overset{\rightarrow}{\partial_p}-\overset{\leftarrow}{\partial_p}\overset{\rightarrow}{\partial_q}\big)g\equiv -f\Lambda g,
\ee
 where $j$ runs through all the degrees of freedom (we will focus on one-dimensional systems hereafter), and arrows indicate the direction of action of the differential operators. The Poisson bracket
governs the dynamics of observables via the canonical equations of motion $\dot q=\{q,H\}$ and $\dot p=\{p,H\}$,
that are generated by the system's Hamiltonian $H$.

Our knowledge about a system is encapsulated in a probability (Liouville) density $\rho(z)$. Its evolution is given by the Liouville equation,
\begin{equation}
\pdv{\rho(q,p)}{t}=\{H,\rho\}=-H\Lambda \rho.\label{Leq}
\end{equation}

The most common {\em quantum} phase space representation~\cite{984WigRev,995PhaDis,ZFC:05} is the Wigner function $W(q,p)$, a real function which may be negative for regions of $q,p$, and is therefore termed a quasi-probability density~\cite{ZFC:05,L:17,932WigFun,977WigPar,984WigRev,000QuAcAn}. The Wigner function corresponding to a Hermitian operator $\hat A$ is the Fourier transform of the off-diagonals of $\hat A$ (the Wigner transform of $\hat A$):
\begin{equation}
A(q,p)\defeq\text{Wigner}_k\{\hat A\}(q,p)=\int dx e^{i {px}/{k}}\expval{q-\tfrac{1}{2} x|\hat A|q+\tfrac{1}{2} x}.\label{Wignertrans}
\end{equation}
The Weyl transform \cite{ZFC:05,984WigRev}
\begin{equation}
\hat A= \frac{1}{4\pi^2k^2}\int \text{Wigner}_k\{\hat A\}(q,p)e^{i\frac{a(q-\hat q)+b(p-\hat p)}{k}}dqdpdadb,\label{Weyltrans}
\end{equation}
effects the inverse transformation, with obvious generalization to $n$ degrees of freedom. In quantum mechanics $k=\hbar$. Both transforms do not affect the dimension.  Since $\hat{\rho}$ is dimensionless,  the Wigner function as a quasi-probability distribution on $\mathfrak{P}$ is defined as $W_{\hat \rho} \defeq \text{Wigner}_\hbar\{\hat\rho/(2\pi \hbar)\}$ \cite{984WigRev}.

The Born rule is reproduced by the following inner product,
\begin{equation}
p(i|\hat \rho_j)=\text{Tr}({\hat E_i} \hat \rho_j )=h\int W_iW_jdqdp,\label{WignerBorn}
\end{equation}
where $h=2\pi\hbar$, and $W_i$ and $W_j$ are Wigner functions corresponding to ${\hat E_i}$ (the effect) and $\hat{\rho}_j$ (the state), respectively.

The (non-commutative) product of operators is represented as $\text{Wigner}_\hbar\{\hat A \hat B\}=\text{Wigner}_\hbar\{\hat A\} \star \text{Wigner}_\hbar\{ \hat B\}$ where
$\star=\exp( -\tfrac{1}{2}i\hbar\Lambda)$ is the Moyal star product.
Finally, the time evolution of the density operator $\hat{\rho}$ under the (Weyl-ordered) Hamiltonian $\hat{H}$ is equivalently represented as the evolution of the Wigner function $W_\rho$, 
\begin{equation}
\frac{\partial W_\rho}{\partial t}=\frac{1}{i\hbar}\left(H\star W_\rho-W_\rho\star H\right)=- \frac{2}{\hbar} H(q,p) \sin(\frac{\hbar}{2}\Lambda) W_\rho(q,p).\label{WFEOM}\end{equation}

 When $\hbar\rightarrow 0$, Eq.~\eqref{WFEOM} becomes the classical evolution Eq.~\eqref{Leq}.

\noindent { {\bf {\em Generalised Born rule and inner product from symmetries.---}}}A key ingredient of any generalised probabilistic theory (GPT) is the assignment of probabilities of the outcomes $i$ of tests on preparations $f$, $P(i|f)=e_i(f)$ ~\cite{001FivAxi, 007GPT,014GPT,012RevDua,013UncDyn,011InPoCo}. In the terminology of quantum foundations research, the functional $e_i$ on the state space is called an effect \cite{BGL:95,BLPY:16}. In GPTs the states are represented as real vectors, which here correspond to continuous real distributions $f(z)$~\cite{016TunNeg}.

\color{black}
It is standard to assume linearity of the effects in a GPT, such that the probability of a discrete outcome can always be represented via
\begin{equation}
\label{eq:probigivenf}
P(i|f)=e_i(f)=c_i\int f(z)g_i(z)dz,
\end{equation}
where $g_i$ is a real-valued normalized function and $c_i$ is a positive constant. $g_i$ does not necessarily represent a valid state. 
For the effects to form a complete measurement the identity $\sum_i e_i(f)=1$ should hold for any allowed state $f$. Thus $\sum_i c_i g_i=1$, which is known as the {\em completeness condition} for a measurement.

Similar expressions give the probability of continuous effects labelled by a continuous variable $\mu$. The probability of falling into an interval $(\mu,\mu+d\mu)$ is $dP(\mu;d\mu|f)=\rho(\mu|f)d\mu$, where $\rho(\mu|f)$ is the probability density for the outcome $\mu$ given the state $f$. We can represent the density by the general expression  $\rho(\mu|f)=c_{\mu} \int f(z)g_\mu (z)dz$. For example,  classical (sharp) phase space localization has $\mu=z_0\in\mathfrak{P}$, and the state  is given by the Liouville density: $f=\rho(z)$. The probability of being within the  volume $dz_0$ around $z_0$ in $\mathfrak{P}$ is $dP=\rho(z_0)dz_0$. Comparison with the general expression identifies $g_{z_0}(z)=\delta(z-z_0)$ and $c_{z_0}=1$.

While the $g_i$ of effects in Eq.~\eqref{eq:probigivenf} are in general not associated with specific states, it is possible that $g_i$ is a function representing a state such that Eq.~\eqref{eq:probigivenf} can be interpreted as the probability being proportional to the inner product between two states, with  \be\expval{f,g}=\int f(q,p) g(q,p) dqdp\label{eq:uniqueinnerprod}\ee being a possible form of the inner product. We find that Eq. \eqref{eq:uniqueinnerprod} is, up to a multiplier, the unique inner product under three symmetries. The symmetries read as follows: 1. Translation: $(q,p,t)\mapsto(q+a,p+b,t+c),$ for any $a,b,c\in \mathbbm R$, 2. Switch: $(q,p,t)\mapsto (Cp,q/C,-t)$, where $C$ is an arbitrary constant fixing unit, 3. Time reversal: $(q,p,t)\mapsto (q,-p,-t)$. (See the Supplementary Material for details).

Eq.~\eqref{eq:uniqueinnerprod} allows us to interpret the inner product with a state $g_i$ as a possibly allowed effect $e_i(f) \propto \expval{g_i,f}\propto \int fg_idqdp$. The proportionality constant is fixed if $e_i(g_i)=1$:
\begin{equation}
\label{eq:ci}
1=e_i(g_i)=c_i\int g_ig_idqdp:=c_i\|g_i\|^2,
\end{equation}
resulting in $c_i= \|g_i\|^{-2}$. We call complete sets of such effects \textit{state-dual measurements}. Projective measurements in quantum theory are an example. State-dual measurements are guaranteed to exist in so-called self-dual theories~\cite{014GPT, 007GPT}.

Eq.~\eqref{eq:ci} associates a property of state $g_i$ with the corresponding state-dual measurement. This relationship ascribes a quantity with the units of $[qp]$ to the effect $g_i$. We will show this $c_i$ can be given the meaning of volume that is occupied by the corresponding state in $\mathfrak{P}$.

\noindent { {\bf {\em Generalised Planck constant of uncertainty: state volume.---}}}
Consider a set of state-dual measurements on the system whose states are confined within the region $\mathfrak{D}\subset\mathfrak{P}$.  The functions $\{g_i\}$ that represent the effects have joint support in the same domain.
Completeness of the measurement inside $\mathfrak{D}$ implies $\sum_{i=1}^{N}c_ig_i= 1_\mathfrak{D}$, where  $1_\mathfrak{D}$   takes value 1 inside the domain $\mathfrak{D}$ and  0 outside it.
As the support of any $g_i$ is in $\mathfrak{D}$, the phase space volume $ V_\mathfrak{D}$ satisfies
\be
\label{eq:VD}
 V_\mathfrak{D}=\int 1_{\mathfrak{D}}dz= \sum_ic_i\int_{\mathfrak{D}}g_idz=\sum_ic_i,
\ee
enabling the interpretation of the coefficients  $c_i$ as the effective phase space volume of the states $g_i$.
We will accordingly use the terminology of the \textit{state volume}  $V_{f_i}$ of a function $f_i$ in $\mathfrak{P}$ as
\begin{equation}
\label{eq:statevoldef}
V_{f_i}:=\left(\int f_i^2dqdp\right)^{-1}=\frac{1}{\|f_i\|^2}.
\end{equation}

The generalised Born rule for state-dual measurements can now be written as
\begin{equation}
\label{eq:generalizedBorn}
P(i|f)=V_{g_i}\int g_i(q,p)f(q,p)dqdp. 
\end{equation}

For example, in quantum theory the purity of a state $\hat\rho$  is bounded via Eq.~\eqref{WignerBorn} as $\mathrm{Tr}\big(\hat\rho^2\big)=h\int W_i^2\leqslant 1$. Thus the state volume of any pure quantum state is given by the Planck constant, {$V_\rho=\|W_\rho\|^{-2}=h$}, while mixed states have larger state volumes.

Classical pure states are associated with points in $\mathfrak{P}$ and Dirac-delta distributions centred on those points \cite{L:17}. For concreteness,   consider a mixed state $f_{\delta\epsilon}$ that is given by a uniform distribution in the volume $\Delta q\Delta p$ where we set $\Delta q=\delta$, $\Delta p=\epsilon$ and take the limit of zero uncertainty by $\epsilon\delta\to 0$. 
A normalised rectangular function is $1/(\epsilon\delta)$ on this domain and zero elsewhere.  Eq.~\eqref{eq:statevoldef} then implies that in the limit $\epsilon\delta\to 0$
the volume $V_{f_{\delta\epsilon}}$ approaches zero. In an epistemically-restricted classical theory simulating quantum mechanics \cite{BST:12} $V_f\geqslant h$.

If we demand that similarly to classical and quantum theories  all pure states in a GPT have the same 2-norm (this does not hold in the example of the probabilistic theory known as box-world \cite{gross2010all}) we can define a state-independent generalization of $h$ as
$\|g_p\|^{-2}=V_p, $
where $g_p$ is an arbitrary pure state. The number $N$ of distinguishable states (associated with the state-dual measurements in $\mathfrak{D}$) can be interpreted as the amount of information (as measured in the number of states) one can store in the system, and then obeys  $N\leq\frac{V_\mathfrak{D}}{V_p}$ in line with Eq.~\ref{eq:VD}.

\noindent {{\bf {\em Generalised energy eigenstates.---}}}\label{Pss}We generalise the energy eigenstates of quantum mechanics as the set of purest stationary states of a GPT. For stationary states the probabilities of all time-independent effects are time-independent, and thus they are given by time-independent functions on $\gP$.

Probabilistic mixtures of stationary states are, by inspection, also stationary, so there is a convex set of stationary states. {\em Pure stationary states} are the extreme points of the set of stationary states.  Pure stationary states are not necessarily pure states of the corresponding GPT, i.e.\ the extreme points \cite{L:17,BLPY:16} of the convex set of all states.

Wigner functions that represent the energy eigenstates of quantum mechanics are stationary by construction. On the other hand, if the action-angle $I-\theta$ variables can be introduced~\cite{Arnold,AKN:10,Z:18}, then the invariance of the action variables is an explicit manifestation of stationarity. The classical energy eigenstates are then $\delta(I-I_0)/(2\pi)$, for all the possible $I_0$, corresponding to uniform distributions over phase-space orbits. Thus pure stationary states of classical mechanics are not classical pure states. These correspond to the phase space points and in the Schr\"odinger picture are explicitly given as $f_{z_0}(t)=\delta\big(z-z_0(t)\big)$, where $z_0(t)\in\gP$ is the phase space trajectory. The pure stationary states moreover coincide with the eigenfunctions of the Liouvillian operator in the Koopman--von Neumann quantum-like formulation of classical mechanics~\cite{longversion}. However, the dual role of the Hamiltonians as the generator of dynamics and as an observable, which we incorporate in the GPT framework, is not respected in the Koopman--von Neumann formulation, which has important consequences for the hybrid quantum-classical mechanics~\cite{023HybMod}.

We will show in subsequent sections that pure stationary states further satisfy two natural desiderata for generalised energy eigenstates: (i) pure stationary states can be assigned sharp energy values, always giving the same value in an energy measurement, and  (ii) they determine the time evolution of the system.

\noindent {{\bf {\em Generalised equation of motion.---}}}A class of generalised equations of motion for the states $f(z,t)$ is obtained if their generator $\mathcal{G}$ is assumed to be a bilinear functional of the state $f$  and the generalised energy eigenstates $\mathcal{G}(f,\sum e_i g_i)=\sum_i e_i\mathcal{G}(f,g_i)$.
Imposing the additional assumptions of (i) the symmetries of canonical coordinates; (ii) preservation of the inner product; (iii) $\mathcal{G}(g_i,g_j)=0$ for all $i,j$ results in (See Supplementary Material for details):
\begin{align}
\pdv{f}{t}
&=\sum_i \epsilon_i \int \mathcal{K}(k)  f \sin(\frac{k}{2}\Lambda) g_i dk \notag \\
&=\sum_i \epsilon_i \int  \frac{i}{2} \mathcal{K}(k) \text{Wigner}_{ k}\{[\hat f_{ k},(\hat{g}_i)_{k}] \} d k,  \label{Commutator EOM}
\end{align}
where $\epsilon_i$ are constant coefficients and  $\mathcal{K}(k)$ is a theory-specific distribution.  $\text{Wigner}_k\{~\}$ represents the Wigner transform of Eq.~\eqref{Wignertrans}, and $\hat f_k, (\hat{g}_i)_{k}$ are the Weyl transforms (Eq.~\eqref{Weyltrans}) of $f$ and $g_i$ (their units are different from density matrices). To include the continuous spectrum (unbounded quantum states, classical mechanics), the sum over should be replaced by integration. This generalised evolution provides a restricted version of the generalised Moyal bracket in Ref.~\cite{022ToogeneralEOM}, here derived from physical principles.

We recover the quantum dynamics (Eq.~\eqref{WFEOM})  if we identify  $\mathcal{K}(k)=2\delta(k-\hbar)/k$ (and $H(q,p)\equiv \sum_i\epsilon_i g_i$). To obtain the classical theory (Eq.~\eqref{Leq}) we have to take a (singular) limit $\hbar\to 0$. Thus $\mathcal{K}(k)$ can be viewed as generalising the {\em dynamically} important Planck constant.

As a simple example of a self-consistent theory where the dynamical and state Planck constants differ, consider d-dimensional quantum systems with a restriction on the information about the preparation such that the allowed pure states of the restricted theory are the states of the form $\frac{1}{2}\ket{\psi}^{(a)}\bra{\psi}^{(a)}+\frac{1}{2}\ket{\psi}^{(b)}\bra{\psi}^{(b)}$ where $\ket{\psi}^{(x)}=\sum_{i=1}^d c_i^{(x)} \ket{i}$ and $\bra{\psi}^{(a)}\ket{\psi}^{(b)}=0$. Then the generalised Planck constant for states is $V=2h$, whereas the dynamical Planck constant remains $h$ (see Ref. \cite{longversion} for details).

 For a general $\mathcal{K}(k)$ the evolution is given by the integral of commutators with different commutation relations. A non-associative algebra replaces the associative Moyal bracket or the operator product. Therefore, the transformation is no longer described by a Lie group, but a quasigroup, or what may be termed a {\em loop} \cite{016QuaGro}. We have not found any principle that rejects this possibility.

\noindent { {\bf {\em Generalised energy and Hamiltonian.---}}}We now complete the discussion by providing an explicit expression for the Hamiltonian as an observable and writing the generalised time evolution equation in terms of that Hamiltonian.

The set of energy eigenstates provides a set of state-dual measurement effects in quantum mechanics. Generalizing this idea, we postulate that there exists a state-dual measurement corresponding to the pure stationary states $\{g_i\}$ ($\{g_{\mu}\}$ in the case of a continuum labelled by $\mu$).

A restriction on how to define generalised energy values is that the identification    $H=\sum_i \epsilon_i g_i$  in Eq.~\eqref{Commutator EOM} in the quantum cases indicates that the energy eigenvalues (scalars with the dimension of energy) are $E_i=\epsilon_i/(2\pi \hbar)=\epsilon_i/ V_i$, where we used Eq. \eqref{eq:generalizedBorn} and that $V_i=2\pi \hbar$ for all pure states in quantum mechanics.

Extending the definition of the Hamiltonian as the generator of time evolution, we demand
\begin{equation}\label{geom2}
\frac{\partial f}{\partial t}=
\int dk \mathcal{K}(k) f\sin(\frac{\Lambda k}{2})H,
\end{equation}
i.e. the dynamics of a GPT is determined by the set $(g_i,\epsilon_i)$ and the kernel $\mathcal{K}(k)$. Thus the Hamiltonian can be written as
\begin{equation}
\label{eq:HamiltonianDef}
H(q,p)=\sum_i E_i g_i V_{g_i}+\int E_\mu g_\mu dV_\mu,
\end{equation}
an expression which also defines the generalised energy eigenvalues (both the discrete and continuous parts of the spectrum). In Ref.~\cite{longversion}, we give a detailed discussion about how the generalised energy is a conserved and additive quantity.

We have seen that the above definition reduces to the standard expression for the energy in quantum theory. In classical mechanics we have $g_{I_0}(I)=\frac{1}{2\pi}\delta(I-I_0)$ where $I$ stands for the action in the action-angle coordinate~\cite{Arnold}, such that
\be
H=\int E_{I_0} \frac{1}{2\pi}\delta(I-I_0) 2\pi dI_0,
\ee
where $E_{I_0}$ is the classical energy that corresponds to the $I_0$, $\frac{1}{2\pi}\delta(I-I_0)$ is a normalised state, and $2\pi dI_0$ gives $dV_{I_0}$. It trivially satisfies Eq.~\eqref{eq:HamiltonianDef}.

Consider the expectation value of energy.
By the generalised Born rule of Eq.~\eqref{eq:generalizedBorn}, $P(i|f)=\int V_{g_i}g_if dqdp$. Combining that with the definition of the Hamiltonian (Eq. \eqref{eq:HamiltonianDef}), we have that the expectation value of energy for state $f$ is given by
\begin{equation}
\expval{E}= \int H fdqdp,
\end{equation}
which for the generalised energy eigenstate $g_i$ is just its value $E_i$.

\begin{table*}
  \centering
 \begin{tabular}{|c|c|c|}
 \hline
Quantum&Classical&Generalization\\\hline
  Wigner function&Non-negative phase space distribution&Arbitrary distribution in the phase space\\\hline
 Energy eigenstates&Delta functions of action&Pure stationary states\\\hline
 uncertainty $h$&0&State volume\\\hline
 dynamical $h$&$\{q,p\}=1$& Non-localized dynamics factor $\mathcal{K}(k)$ in Eq.~\eqref{Commutator EOM} \\\hline
Equation \eqref{WFEOM}&Liouville equation Eq.~\eqref{Leq}& Time evolution Eq. (\ref{geom2})\\
\hline
  \end{tabular}
  \caption{Comparison between quantum, classical and our generalised framework.}\label{compare}
\end{table*}
\noindent {{\bf {\em Summary and outlook.---}}}We built a generalised phase-space framework centred around generalizations of the quantum energy concepts, like Hamiltonian and energy eigenstates (as listed in Table~\ref{compare}). We define the generalised energy eigenstates operationally: the most pure stationary states. Based on these pure stationary states, we derive a generalised equation of motion in phase space which encompasses the quantum and classical Liouville equations of motion. This includes generalizing Planck's constant. In our framework, Plank's constant provides the volume occupied by pure states and also appears in the commutation relation in the equation of motion.  The two generalizations of Planck's constant can have different values in general theories. The axioms used are listed together in the Supplementary Materials.   A specific theory is obtained by specifying the set of pure states of the theory, the dynamical kernel $\mathcal{K}(k)$, and a general post-measurement state update rule, as can be seen e.g.\ from comparison with summaries of classical and quantum axioms \cite{L:17,006QuCoMe,BLPY:16}. An accompanying extended article contains a derivation of the generalised Born rule from symmetries and examples of theories other than quantum and classical, amongst other things~\cite{longversion}.

This framework can be employed and developed in several directions: (i) the link between the generalised evolution, state/effect negativity, `jumping in phase space' and contextuality deserves investigation~\cite{longversion}.
(ii) other forms of mechanics can be built, that are neither classical nor quantum, e.g. by letting Plank's constant in the equation of motion differ from Plank's constant for uncertainty or choosing a non-trivial $\mathcal{K}(k)$, (iii) the framework enables clear analogies and comparisons between quantum and classical dynamics and could be for example help clarify the apparent speed-up of Hamiltonian-based quantum walks over classical walks~\cite{venegas2012quantum}, (iv) it may be possible to reduce or alter the set of postulates (see the Supplementary Material for a list), (v) it is natural to employ the framework to create a theory of thermodynamics that is independent of the underlying choice of mechanics.

\noindent { {\em Acknowledgements.}} We are grateful to Andrew Garner and Dario Egloff for early discussions and to Meng Fei, Fil Simovic, Chelvanniththilan Sivapalan and Ioannis Soranidis for feedback on the draft. LJ and OD acknowledge support from the National Natural Science Foundation of China (Grants No.12050410246, No.1200509, No.12050410245) and City University of Hong Kong (Project No. 9610623). The work of DRT was in part supported by the Shenzhen Institute for Quantum Science and Engineering,
Southern University of Science and Technology

\end{document}


\title{Supplementary Material}
\author{Libo Jiang }
\email{11930020@mail.sustech.edu.cn}
\affiliation{Alumnus,
Southern University of Science and Technology (SUSTech), Shenzhen 518055, China}

\pacs{123 }

\author{Daniel Terno}
\email{daniel.terno@mq.edu.au}
\affiliation{Department of Physics and Astronomy, Macquarie University, Sydney, New South Wales 2109, Australia}
\pacs{23 }

\author{Oscar Dahlsten}
\email{oscar.dahlsten@cityu.edu.hk}
\affiliation{Department of Physics, City University of Hong Kong, Tat Chee Avenue, Kowloon, Hong Kong SAR, China
}%
\affiliation{Shenzhen Institute for Quantum Science and Engineering and Department of Physics,
Southern University of Science and Technology, Shenzhen 518055, China}

\affiliation{Institute of Nanoscience and Applications, Southern University of Science and Technology, Shenzhen 518055, China}
\pacs{12 }
\maketitle
\section{List of postulates}

\begin{assume}[Canonical coordinate symmetries]\label{canon}
There exists a coordinate system $(q,p)$  where the physical laws manifested by equations of motion and measurement are invariant under the following coordinate transformations:\\
 1. Translation: $(q,p,t)\mapsto(q+a,p+b,t+c),$ for any $a,b,c\in R$. {We represent its action on functions via $(\hat{T}_{a,b,c}f)(q,p,t)=f(q+a,p+b,t+c)$}.\\
2. Switch: $(q,p,t)\mapsto (Cp,q/C,-t)$, where $C$ is an arbitrary constant with units $[C]=[ {q}/{p}]$. \\
3. Time reversal: $(q,p,t)\mapsto (q,-p,-t)$. (equivalent to $(q,p,t)\mapsto (-q,p,-t)$ by switch.)
\end{assume}

\begin{assume}[Local inner product]\label{local}
The inner product is local which means for two arbitrary quasi-probability distributions $f_1$ and $f_2$,
  \begin{equation} \lim_{a\rightarrow\infty}\expval{f_1,\hat{T}_{a,0,0}f_2}=0.\end{equation}
\end{assume}

\begin{assume}[Evolution dependence]\label{PSSEES}
The time evolution of a state only linearly depends on the \textit{pure stationary states}, up to some dimensional factors $\mathcal{E}_i$ to keep the dimensions identical.
\begin{equation}
\label{eq:PSSEES}
\frac{\partial f}{\partial t}=G\left(f,\sum_i \mathcal{E}_i g_i \right)=\sum_i \mathcal{E}_i G\left(f, g_i \right), \end{equation} where $g_i$ is a set of pure stationary states and $\mathcal{E}_i$ are corresponding parameters and $G$ is some bi-linear functional.
\end{assume}

\begin{assume} [Independence of stationary states]
The pure stationary states are independent in the sense that $G\left(g_i, g_j\right)=0$ holds for arbitrary $i,j$.\label{pssees}
\end{assume}

\begin{assume}[Inner product invariance]\label{infcon}
The time derivative of inner products $\frac{\partial}{\partial t}\int f_1(t)f_2(t)dqdp=0$ for arbitrary states $f_1,f_2$ and time point $t$.\end{assume}

\begin{assume}[Existence of energy measurement]There exists a state-dual measurement whose effects all correspond to pure stationary states. \label{measure}
\end{assume}

\section{the inner product from canonical symmetries }
This derivation depends on Postulate \ref{canon} and \ref{local}.

An inner product is a bilinear symmetric function of two states. For phase space distributions $f_1$ and $f_2$, a general form of such a bilinear function is
\begin{equation}
\label{eq:GeneralInnerProd}
\int M(q,p,\Delta q,\Delta p)f_1(q,p)f_2(q+\Delta q,p+\Delta p)dqdpd\Delta qd \Delta p,
\end{equation}where $M$ is an arbitrary function. The symmetric condition on the  inner product $\expval{f_1,f_2}=\expval{f_2,f_1}$ further requires
\begin{equation}
\label{eq:a}
M(q,p,\Delta q,\Delta p)=M(q,p,-\Delta q,-\Delta p)
\end{equation}
 for arbitrary $a,b,c,d\in \mathbbm{R}.$

Translation symmetry requires $\expval{f_1(q,p),f_2(q,p)}=\expval{f_1(q+a,p+b),f_2(q+a,p+b)}$ such that
\begin{equation}\begin{array}{l}\label{eq:translationM}
\int M(q,p,\Delta q,\Delta p)f_1(q,p)f_2(q+\Delta q,p+\Delta p)d\Omega=\\
\int M(q,p,\Delta q,\Delta p)f_1(q+a,p+b)f_2(q+a+\Delta q,p+b+\Delta p)d\Omega,
\end{array}
\end{equation}where $d\Omega=dqdpd\Delta q d \Delta p$.
Eq. \eqref{eq:translationM} holds for arbitrary $f_1$, $f_2$, so
\begin{equation}
M(q,p,\Delta q,\Delta p)= M(q-a,p-b,\Delta q,\Delta p),
\end{equation}
for all $a,b\in \mathbbm{R}$. Therefore, $M$ only depends on the relative distance $\Delta q, \Delta p,$.
\begin{equation}
\label{eq:b}
M(q,p,\Delta q,\Delta p)=M(\Delta q,\Delta p).
\end{equation}
Similarly, switch symmetry with dimensional constant $C$ requires $\expval{f_1(q,p),f_2(q,p)}=\expval{f_1( p/C,C q),f_2(p/C,C q)}$, which leads to
 \begin{equation}
 \label{eq:c}
M(\Delta q,\Delta p)=M(\Delta p/C,C\Delta q).
\end{equation}
Time reversal symmetry requires $\expval{f_1(q,p),f_2(q,p)}=\expval{f_1(q,-p),f_2(q,-p)}$, which leads to
 \begin{equation}
 \label{eq:d}
M(\Delta q,\Delta p)=M(\Delta q,-\Delta p).
\end{equation}

Equations~\eqref{eq:a}, \eqref{eq:b}, \eqref{eq:c}, \eqref{eq:d} imply that $M(q,p,\Delta q,\Delta p)$ is constant when $|\Delta p\Delta q|=c$ for arbitrary $c\geq 0$, except at the origin $(\Delta q=\Delta p=0)$. All these contour lines extend to infinity. Nevertheless, we postulated the local inner product, $\langle f_1, f_2 \rangle=0$ for infinitely separated states, $M(\Delta q,\Delta p)$ must go to 0 when $\Delta q \rightarrow \infty$. This implies that $M(\Delta q,\Delta p)=0$ except for at the origin ($\Delta q=\Delta p=0$). Thus, $M(\Delta q, \Delta p)\propto \delta(\Delta q)\delta(\Delta p)$, and the inner product must have the form
\begin{equation}
\expval{f_1,f_2}\propto\int f_1(q,p)f_2(q,p)dqdp.
\end{equation}

\section{Deriving the equation of motion}
This is an abbreviated derivation depending on Postulate \ref{canon}, \ref{PSSEES}, \ref{pssees}, and \ref{infcon}. The aim is to get the generalized equation of motion:
\begin{equation}
\pdv{f}{t}=\sum_i \epsilon_i \int \mathcal{K}(k)  f \sin(\frac{k}{2}\Lambda) g_i dk.\label{eq1}
\end{equation} In this equation, $\pdv{f}{t}$ only depends on local derivatives of $f$ and generalized eigenstates $g_i$. To derive it, we introduce an equivalent but different form of the equation:
\begin{equation}
\frac{\partial f}{\partial t}(q,p)=\int f(q+l,p+j)J(q,p,l,j)dld.\label{eq2}
\end{equation}
When $J(q,p,l,j)$\begin{equation}
  =\sum_i \epsilon_i \text{Im}\int g_i(q+y,p+z) A(k')  e^{-ik'(jy-lz)} dk'dydzdldj,  \label{eq3}
\end{equation} the Eq. \eqref{eq2} is equivalent to Eq. \eqref{eq1}, where $A(k')$ corresponds to the $\mathcal{K}(k)$, they represent the exactly same degree of freedom. We can find the $\frac{\partial f}{\partial t}(q,p)$ not only depends on the functions at $(q,p)$, but the whole phase space ($dldj$ are integrated over the whole phase space). You can check the two different forms are equivalent by multivariate Taylor expansions of $f(q+l,p+j)$ and $g(q+y,p+z)$ in Eq. \eqref{eq2}, \eqref{eq3}.

The Eq. $\eqref{eq2}$ describes the most general evolution of quasi-probability. We are going to restrict the $J(q,p,l,j)$ by postulates. The inner product invariance $\int\frac{\partial f_1}{\partial t}f_2+\frac{\partial f_2}{\partial t}f_1dqdp=0$ requires \begin{equation}J(q,p,l,j)=-J(q+l,p+j,-j,-l).\label{cond1}\end{equation}

We have introduced symmetries of canonical coordinates: Switch, Time reversal,  Translation. Inspired by quantum mechanics, we assume the evolution linearly depends on the generalised energy eigenstates $g_i$, so $J$ is a linear functional of $g_i$.  When we apply these symmetry operations to state $f$ as well as eigenstates $g_i$, we expect the equation of motion to still hold after operations:
\begin{equation}\pdv{f'}{t'} (q,p)=\int {J}_{g'}(q,p,l,j) f'(q+l,p+j)dldj.\label{exofsym}\end{equation}
(We can write $(q,p,t)\mapsto(q',p',t')$, or equivalently $f(q,p),g(q,p),t\mapsto f'(q,p),g'(q,p),t'$. We choose the latter one here.)
The switch and time reversal symmetries require $J(0,0,j,l)=J(0,0,-j,-l)$. Adding translation symmetry we require \begin{equation}
    J(q,p,j,l)=J(q,p,-j,-l)\label{cond2}
\end{equation} for arbitrary $q,p,l,j$.

Combining Eq. $\eqref{cond1}, \eqref{cond2}$, we find  \begin{equation}\label{period} J(q,p,l,j)=J(q+l,p+j,l,j),\end{equation} which means $J$ is periodic on $q,p$. Hence, only certain frequency components satisfying $k_ql+k_pj=2\pi n~(n\in Z)$ are allowed in $J$.
A general linear functional of $g$ is given by
\begin{equation}
J(q,p,l,j)=\text{Re}\int \tilde g(k_q,k_p) A'''(k_q,k_p,l,j)e^{i(k_qq+k_pp)}dk_qdk_p,
\end{equation}  where $A'''$ and all the $A$s below represent some unsettled degrees of freedom in $J$, different $A$s help to absorb constants. The periodic property adds a $\sum_n\delta(k_ql+k_pj-2\pi n)$ term in the frequency domain.
\small
\begin{equation}\begin{array}{l}\label{eq:fourierJ}
J(q,p,l,j)\\
=\text{Re}\int \tilde g(q_k,p_k) A''(k_q,k_p,l,j)\sum_n \delta(k_ql+k_pj-2\pi n)e^{i(k_qq+k_pp)}dk_qdk_p\\
=\text{Re}\int g(q+y,p+z) \sum_n A'\left(\frac{2\pi n-k_pj}{l},k_p,l,j\right) e^{i(\frac{2\pi n-k_pj}{l}y+k_pz)}dk_pdydz\\
=\text{Re}\int g(q+y,p+z) \sum_n A(n,k',l,j) e^{i\frac{2\pi ny}{l}} e^{-ik(jy-lz)} dk'dydz,
\end{array}\end{equation}
\normalsize
where we relabelled $\frac{k_p}{l}$ by $k'$. The term $e^{i\frac{2\pi ny}{l}}$ is not well-defined when $l=0$, but this term will vanish later.

The requirement Eq. \eqref{cond2} further requires that $J$ is an odd function of $l,j$, so
\begin{equation}J_{g}= \text{Im} \int g_i(q+y,p+z) \sum_n A(n,k',l,j)  e^{i\frac{2\pi ny}{l}} e^{-ik'(jy-lz)} dkdydz,\end{equation}and requires $A(n,k',l,j)=A^*(n,k',-l,-j)$.

One more requirement is the pure stationary states themselves should be stationary under the equation of motion (Postulate \ref{pssees}), i.e.,
\begin{equation}\label{eq:kprimeconstraint}
\begin{array}{l}
~~~~\int g(q+l,p+j)J_g(q,p,l,j)dldj \\
=\text{Im}\int g(q+l,p+j) g(q+y,p+z)\\ ~~~~~~~~~~~~~~~~~~~~~~~~~~~\sum_n A(n,k',l,j)  e^{i\frac{2\pi ny}{l}} e^{-ik'(jy-lz)} dk'dydzdldj\\=0.
\end{array}
\end{equation}
Observe that the equation can be written in the matrix form: \begin{equation}g_{lj}M^{ljyz}g_{yz}=0,\end{equation} where $M^{ljyz}=\text{Im}\int\sum_n A(n,k',l,j)  e^{i\frac{2\pi ny}{l}} e^{-ik'(jy-lz)}dk'$.
 It means $M$ must be a generator of the orthogonal group, which is anti-symmetric, $M^{ljyz}=-M^{yzlj}$, swapping $yz$ with $lz$ changes its sign. Therefore, we require $n$ can only equal zero and $A(n,k',l,j)=A(k')$.
Now the form of $J$ is
\begin{equation}
J_g(q,p,l,j)=\text{Im}\int g(q+y,p+z) A(k')  e^{-ik'(jy-lz)} dk'dydzdldj.
\end{equation} This is exactly Eq. \eqref{eq3}, which means we have derived the generalized equation of motion Eq.~\eqref{eq1}.